# A New Interpretation of Three-D/imensional Particle Geometry: M-A-V-L

Yu Feng Su, Sumana Bhattacharya, and Seung Jae Lee[*]

Department of Civil and Environmental Engineering, Florida International University
Miami, Florida 33174, U.S.A.

Chang Hoon Lee, and Moochul Shin
Department of Civil and Environmental Engineering, Western New England University
Springfield, Massachusetts 01119, U.S.A.

## ABSTRACT

This study provides a new interpretation of 3D particle geometry that unravels the 'interrelation' of the four geometry parameters, i.e., morphology M, surface area A, volume V, and size L, for which a new formula, $M = A/V \times L/6$, is introduced to translate the 3D particle morphology as a function of surface area, volume, and size. The $A/V \times L$ of a sphere is invariably 6, which is placed in the denominator of the formula, and therefore M indicates a relative morphological irregularity compared to the sphere. The minimum possible value of M is clearly one, and M may range approximately to three for coarse-grained mineral particles. Furthermore, the proposed formula, $M = A/V \times L/6$, enables to graphically preserve the four parameters' relations when plotting the geometry parameter distributions. This study demonstrates the approach with two plot spaces that represent (i) L vs. M and (ii) A/V vs. V, where A/V works as the messenger between these two spaces as $A/V = M/L \times 6$. Therefore, this approach helps comprehensively address the four-dimensional aspects of the 3D particle geometry and better understand the parameters' combined influence on the mechanical behavior of granular materials.

---

[*] Corresponding author. Phone: +1-305-348-1086, Email: *sjlee@fiu.edu*

# 1  INTRODUCTION

Granular materials are prevalent in nature such as soil, and important in many industries including construction, agriculture, and others, which are known as the second-most manipulated material in industry next to water [1]. The influence of particle geometry is a key to understand the complex behavior of the granular materials, but our knowledge of the subject remains at best incomplete.

The 3D particle geometry is characterized in terms of four parameters, i.e., morphology, surface area, volume, and size. The quantities of these parameters are interrelated, e.g., if size changes, surface area and volume also change. Morphology is also closely related to size, as morphology affects the size measurement [2] as well as surface area and/or volume. For example, Figure 1a shows two different particle morphology with the same size$^\dagger$. These particles necessarily have different volumes and surface areas from each other. On the other hand, if the particle volume is same as shown in Figure 1b, these particles have different sizes and surface areas as well as the different morphology. If two particles with the same morphology are of different sizes, these particles have different volumes and surface areas (Figure 1c).

The definitions of particle volume, surface area, and size are clear, and each parameter is characterized by a single scalar value. On the other hand, three factors have been traditionally used for the characterization of morphology, which are defined at three different scales: (i) global form (at large scale), (ii) local angularity (at intermediate scale), and (iii) surface texture (at small scale) as shown in Figure 2. The (i) global form provides the largest scale morphological information related to the particle's diameter scale $O(d)$ and characterizes the extent to how the morphology is equidimensional, e.g., sphere vs. ellipsoid. The (ii) local angularity describes the overall sharpness of corners defined at a smaller length scale by one order of magnitude $O(d/10)$ [3,4]. Sphericity and Roundness [5] are the broadly adopted descriptors to characterize the global form and the local angularity. Both Sphericity and Roundness range between 0 and 1. A low Sphericity indicates an elongated shape such as an ellipsoid, while a high Sphericity close to 1 indicates a near-equidimensional shape such as a sphere; a low Roundness indicates a particle with sharp corners, while a high Roundness indicates the opposite. In addition, Regularity $\rho$ was defined as the arithmetic average of Sphericity and Roundness [3]. Compared to the global form and the local angularity that have been optically characterized using Sphericity and Roundness, the effect of (iii) surface texture can be characterized by the inter-particle friction angle [6–8] which has been modeled as an input parameter in the discrete element analysis [9,10].

---

$^\dagger$ This study measures size in terms of the diameter of particle's bounding sphere and will be consistently used throughout the paper.

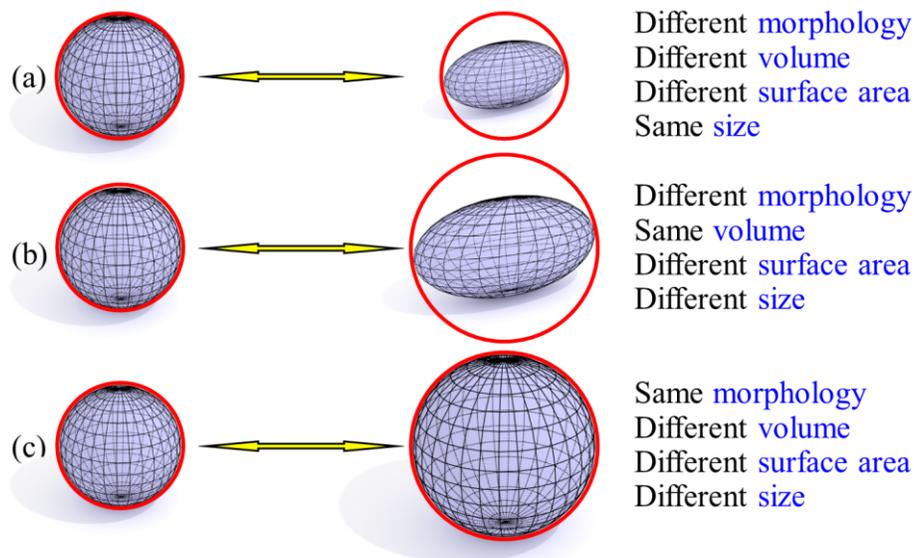

Figure 1. Interrelated 3D particle geometry parameters; (a) Two different particle morphology with the same size, consequently the particles have different volumes and surface areas. The size is measured in terms of the diameter of particle's bounding sphere (shown as red circles); (b) Two different particle morphology with the same volume, thus the particles have different sizes and surface areas; (c) Two different particles with same morphology having different volumes, surface areas, and sizes.

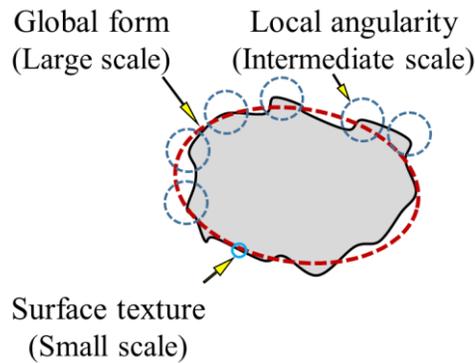

Figure 2. Conventional morphology characterization at three different scales [11].

Sphericity and Roundness are traditionally defined in 2D using the particle projection images [12]. These 2D descriptors have been commonly adopted due to the ease of characterization of global form and local angularity, but which have inherent limitation to interrelating the morphology with volume and surface area that are defined in 3D.

The 'true' Sphericity is a 3D descriptor developed by Wadell [5] that characterizes the particle morphology by comparing its surface area to that of a sphere with the same volume. It is worthwhile to note that the true Sphericity is not an indicator of the global form only, because it is a function of the surface area, thereby also influenced by the local angularity [13,14]. This 3D Sphericity has not been popularly used due to the challenge of measuring the surface area until recently the 3D optical characterization techniques became available in the research community such as X-ray computed tomography [15,16], 3D laser scanning [17,18], photogrammetry [19]. Other 3D Sphericity descriptors also have been developed with the advances in the 3D optical characterization techniques, e.g., Alshibli et al. [20] used the 3D synchrotron microcomputed tomography to obtain 3D images of various soil particles and proposed a new Sphericity index $I_{sph}$ = V / V$_{sph}$, where V$_{sph}$ is the volume of a sphere with a diameter equals to $ds$ which is the shortest principal axis length of the particle that passes through the center of particle mass. Kong and Fonseca [21] proposed another volume-based Sphericity index similar to $I_{sph}$, but which has the maximum value of 1. Bullard and Garboczi [22] introduced two 3D Roundness concepts: $R_W$, a 3D analog of Wadell's 2D Roundness, and $R_n$ for a simple 3D angularity measurement that accounts for the angle between the surface position vector and the unit normal vector over the particle surface. Zhao and Wang [14] introduced a 3D Roundness that considers the curvatures of all particle corners, which may be viewed as another 3D interpretation of Wadell's 2D Roundness. Cruz-Matías and others [23] introduced a 3D Roundness index that considers the difference between a particle and its reference ellipsoid. Although the developed 3D descriptors contributed to a realistic measurement of the 3D particle morphology, the interrelation of morphology to the other 3D geometry parameters has not been considered.

---

**Problem statement #1**: Morphology is closely related to surface area, volume, and size, as demonstrated with the examples in Figure 1, which therefore may be considered as a set of 'morphology parameters.' If so, can we represent morphology M as a function of surface area A, volume V, and size L instead of the conventional approach that characterizes the morphology independently of surface area, volume, and size? While the authors appreciate the importance of characterizing the individual contributions of Sphericity and Roundness to overall morphology, this study aims to address the new question as it would provide a new perspective that unravels the 'interrelation' of the four geometry parameters, which could not be explained with the conventional approach.

---

Previous research mostly considered particle morphology and size as the major parameters that could influence the granular material behavior. Therefore, the size was controlled in the considered granular material specimens to study the morphology effect or the other way around. However, it would be important to comprehensively address the four-dimensional aspects (i.e., volume, surface area as well as morphology and size) of the 3D particle geometry to better understand the particle geometry effect and enhance the predictive capability. For example, the two particle models in Figure 1a having different morphology are of the same size. If these particle models are adopted in particle-based simulations (e.g., discrete element analysis) to study the morphology effect on the granular material behavior, the different volume and surface area may produce additional or uncontrolled effects on top of the effect by different morphology. While previous studies on the granular materials mostly focused on particle morphology or size effect, the volume of solid particles is also an important factor that needs to be systematically considered as it is a major parameter that determines the granular skeletal density. Furthermore, the particle surface area is another important factor, e.g., soil plasticity is greatly influenced by the surface area [24], and also plays a critical role in the performance of cemented granular materials (e.g., concrete) due to its direct impact on the quantity of bonding characteristics between particle surface and binding matrix [25–29]. The previous studies commonly corroborated that the interface of particles and binding matrix, known as the "interface transition zone (ITZ)," plays a critical role in the load-transferring mechanism in the cemented granular materials and the quantity of bonding characteristics at this ITZ is governed by the particle surface area. The authors in elsewhere [29] demonstrated the particle surface area is a critical mediator correlating the particle morphology with the mechanical performance of cemented granular materials due to its direct impact on the bonding quantity at the ITZ.

---

**Problem statement #2**: Previous research has adopted particles of different morphology while controlling the size to study the particle morphology effect on the granular material behavior. However, considering these particles also present different volumes and surface areas, how can we conclude the different behavior is caused by the different morphology only? Hypothetically speaking, different particle morphologies having same volume, surface area, and size need to be considered to understand the morphology effect. However, it is implausible to have such particles as implied by the examples in Figure 1 and unconvincing to control all four geometry parameters to develop such particle models for discrete element analysis. Furthermore, particle

---

volume and surface area are also significant factors that affect the granular material behavior. Therefore, it would be important to comprehensively address the four-dimensional aspects of 3D particle geometry to understand the parameters' combined influence on the granular material behavior as it is hard to independently discuss the effect of one parameter from the others.

Comprehensive characterization of the particle geometry 'distributions' is another important key to understand the causality between the particle geometry effect and the granular material behavior. Previous research mostly focused on the morphology and size effect, in which at least three parameters are characterized, i.e., Sphericity, Roundness, and size. A 3D scatter plot could be developed with data markers to describe the distributions, i.e., each marker representing the geometry of a particle in the 3D plot space, where each axis refers to Sphericity, Roundness, and size. However, to the best of the authors' knowledge, this approach has not been adopted in the engineering practice possibly because a 3D scatter plot is difficult to interpret if looked at one viewpoint and is not immediately obvious. Typically multiple figures are required from different viewpoints to better interpret the 3D scatter plot. Engineers, instead, have described the distributions separately in 2D plot spaces to facilitate the data interpretation [30–35] such as Figure 3. However, the geometry parameters are not cross-referenced with this approach, thus only partial information is deliverable. For example, as illustrated in Figure 3, it is not possible to interrelate Sphericity or Roundness with a particle size of interest, and vice versa.

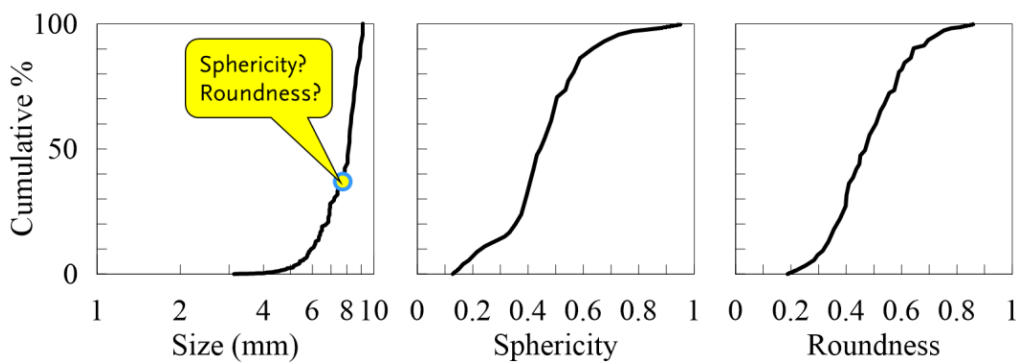

Figure 3. Separate description of particle size and morphology distributions.

**Problem statement #3**: As discussed above, it is already challenging to comprehensively describe the particles' geometry distributions even for three parameters without losing the valuable information regarding the parameters' interrelations. With the need of addressing the four-dimensional aspects of 3D particle geometry enunciated with the problem statements #1 and #2, such comprehensive description will be more challenging with the four geometry parameters. This study aims to introduce a new approach that can graphically preserve the four parameters' relations when plotting in 2D plot spaces, which helps comprehensively describe the four-dimensional aspects of 3D particle geometry distributions.

To address the problem statement #1, this study introduces a new formula in Section 2 that can systematically quantify the interrelation of four 3D particle geometry parameters, morphology M, surface area A, volume V, and size L. With the new formulation, a parameter can be represented as a function of the other three, which also enables to comprehensively address the four-dimensional aspects of the particles' geometry distributions on two 2D plot spaces. This plotting approach to address the problem statement #3 is discussed in Section 3. The new formula and plotting method will help better understand the geometry parameters' combined influence on the behavior of granular materials (the problem statement #2), which is demonstrated in Section 4.

**Note**: While the characterization of surface texture is also of great importance, this study limits its scope on the morphology at the large (i.e. $O(d)$) and intermediate (i.e., $O(d/10)$) scales, because the small-scale surface texture would have relatively insignificant interrelation with volume, surface area, and size. For example, two identically shaped particles except the surface texture can be reasonably assumed to have the same volume, surface area, and size.

## 2 PROPOSED FORMULA FOR THE NEW INTERPRETATION OF THREE-DIMENSIONAL PARTICLE GEOMETRY

This study leverages the fundamental geometry principle: 'morphology is related to surface-area-to-volume (A/V) ratio.' An example is shown in Figure 4, where the A/V ratios are compared for the five different particle models having the same unit volume. The sphere has the smallest A/V ratio, which increases with more angular morphology. The A/V ratio of a cube is 1.24 times higher than that of a sphere with the same volume. The A/V ratio of a regular tetrahedron (which is more angular than the cube) is 1.49 times higher than the sphere. The A/V ratio also increases with elongation. The A/V ratio of a stretched tetrahedron in the figure is 1.65 times higher than the A/V ratio of the sphere with the same volume. With extreme angularity, the A/V ratio becomes significantly larger than that of the sphere as shown for the great stellated dodecahedron in the figure.

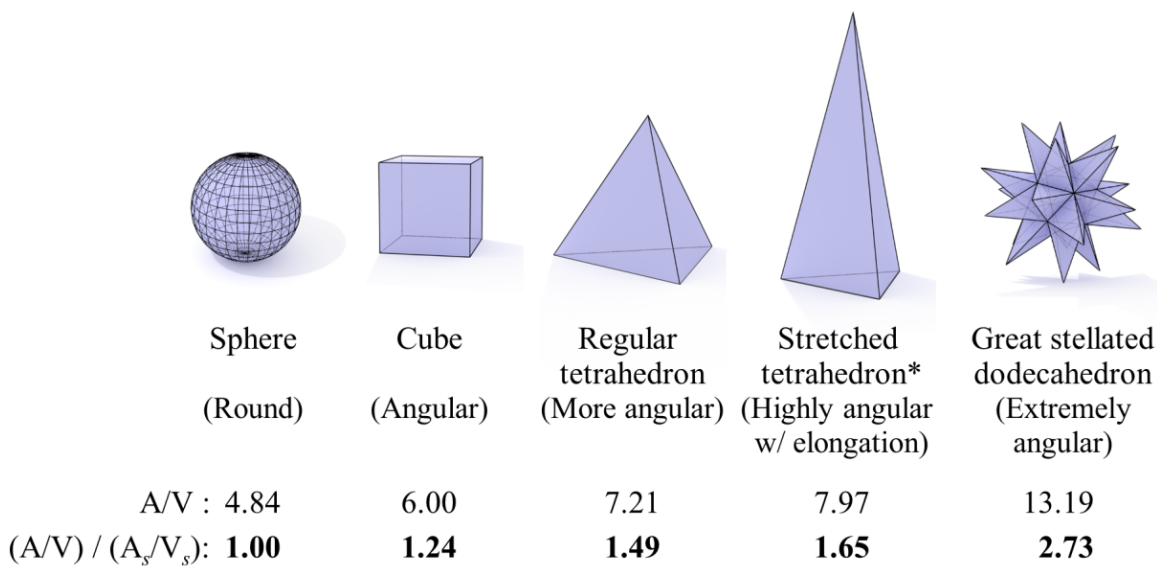

| | Sphere (Round) | Cube (Angular) | Regular tetrahedron (More angular) | Stretched tetrahedron* (Highly angular w/ elongation) | Great stellated dodecahedron (Extremely angular) |
|---|---|---|---|---|---|
| A/V : | 4.84 | 6.00 | 7.21 | 7.97 | 13.19 |
| (A/V) / (A_s/V_s): | **1.00** | **1.24** | **1.49** | **1.65** | **2.73** |

Figure 4. Morphology related to surface-area-to-volume (A/V) ratio; All particles have the same volumes of a unity; The subscript $s$ represents sphere, i.e., $A_s$ and $V_s$ are surface area and volume of the sphere. Therefore, $(A/V) / (A_s/V_s)$ indicates the A/V ratio of a particle relative to the A/V ratio of the sphere with the same volume; *The length of three edges of the stretched tetrahedron is 3.2 each, and the length of the other edges is 1.5.

While the A/V ratio is related to particle morphology, it is important to note that the ratio is not constant for a given morphology because the value depends on the size as implied by its unit, which is a reciprocal length ($L^2/L^3 = L^{-1}$). Therefore, the A/V ratio is inversely proportional to the size. For this reason, the product of A/V ratio and size turns out to be invariant for a given morphology, which therefore may be used as a morphology index. Building upon this concept, this study proposes a new formula as below that interprets the 3D particle morphology M as a function of the other geometry parameters, i.e., surface area A, volume V, and size L:

$$M = (A/V \times L) / (A_s/V_s \times L_s) = (A/V \times L) / 6$$

where the A/V ratio is multiplied by L for the size-independent characterization of particle morphology, which is unit-less. The subscript $s$ in the equation means sphere. Therefore, M indicates the relative morphological irregularity compared to the sphere. The A/V×L does not change for a given morphology and $A_s/V_s \times L_s$ (of the sphere) is invariably 6. Therefore, the minimum possible value of M is clearly 1, and is higher than 1 for typical mineral grains. It is interesting to note that M is a size-independent parameter, although M is expressed as a function of L (size).

|  | (a) | (b) | (c) | (d) | (e) |
|---|---|---|---|---|---|
| Volume (V): | 1.77 cm³ | 0.53 cm³ | 1.77 cm³ | 0.71 cm³ | 0.53 cm³ |
| Surface area (A): | 8.59 cm² | 3.86 cm² | 7.07 cm² | 3.86 cm² | 3.17 cm² |
| Size (L): | 2.25 cm | 1.50 cm | 1.50 cm | 1.11 cm | 1.00 cm |
| A/V: | **4.85** cm⁻¹ | **7.28** cm⁻¹ | **4.00** cm⁻¹ | **5.41** cm⁻¹ | **6.00** cm⁻¹ |
| A/V×L: | **10.92** | **10.92** | **6.00** | **6.00** | **6.00** |
| M = A/V×L/6: | **1.82** | **1.82** | **1.00** | **1.00** | **1.00** |

Figure 5. 3D morphology (M) translated as A/V×L/6; the size is measured in terms of the diameter of the bounding sphere (shown as red circles).

Figure 5 demonstrates how the new formula is interpreted, for which three different spheres (Figure 5c, d, and e) are scaled to match one of V, A, or L (highlighted in gray) of the irregularly shaped particle in Figure 5b. The 3D particle model of irregular morphology is a polyhedron with 10374 triangular faces in its surface mesh, which is developed from a mineral particle after 3D scanning using a photogrammetry technique introduced in Zhang et al. [36]. A larger particle with the same irregular morphology is also shown in Figure 5a, which is 1.5 times larger than the particle in Figure 5b by size. The volume and surface area of the particles in Figure 5a and b are numerically obtained from the developed 3D polyhedron models. Following observations can be made:

(i) Particle morphology is related to A/V ratio, thus the irregular particle in Figure 5b has a higher A/V ratio compared to the sphere of the same L in Figure 5c, (i.e., 7.28 cm$^{-1}$ vs. 4 cm$^{-1}$);

(ii) Since the A/V ratio is inversely proportional to the size, the A/V ratio increases while L decreases. For example, the A/V ratio of the sphere in Figure 5e is larger than that of Figure 5c by 1.5 times (i.e., 6 cm$^{-1}$ vs. 4 cm$^{-1}$), but L is smaller by 1.5 times (i.e., 1.0 cm vs 1.5 cm). This principle also holds for the irregularly shaped particles (Figure 5a, and b): the A/V ratio increases by 1.5 times from 4.85 to 7.28 cm$^{-1}$ and L decreases by 1.5 times from 2.25 to 1.50 cm;

(iii) Therefore, the A/V×L values remain the same for each morphology regardless of the size, i.e., 10.92 and 6.0;

(iv) The A/V×L for any sphere is invariably 6, thus the minimum possible value of M is clearly 1. The example in Figure 5 demonstrates that the higher the morphological irregularity is, the higher the M value becomes. With the Krumbein and Sloss chart [37] we may infer that the maximum value of M can be about 3 for coarse-grained mineral particles in nature, which will be further discussed in Section 4.

With the new formulation M = A/V×L/6, if one knows A, V and L, then morphology can be evaluated in terms of M. On the other hand, in previous studies on the granular materials, the morphology was separately evaluated in addition to the information of A,V, and L. The examples in Figure 1 can be now understood with the proposed formula, M = A/V×L/6, regarding how the change of a geometry parameter may affect the other parameters. In addition, this formula explains the 'unilateral' relationship between morphology and size: it is clear that the change of M impacts A, V, and L values from the formula, which explains change of morphology may change the size (Figure 1b). However, the change of size does not change the morphology (Figure 1c), i.e., simply scaling of the particle size does not change the morphology: the A/V ratio is linearly proportional to 1/L for a given morphology, so the change of L is canceled out by the change of A/V, which makes M invariant of the size. Therefore, the 'unilateral' relationship between morphology and size can be clearly explained with the new formula.

# 3 FOUR-DIMENSIONAL ASPECTS IN THE 3D GEOMETRY PARAMETER DISTRIBUTIONS

## 3.1 M-A-V-L Approach for Comprehensive Description of the Parameter Distributions

The proposed formula, M = A/V×L/6, helps graphically preserve the relations of the four 3D particle geometry parameters when plotting the parameter distributions. This paper demonstrates the approach with two 2D plot spaces, (i) L vs. M and (ii) A/V vs. V, where A/V works as the messenger between the two spaces as A/V = M/L×6. Therefore, this approach helps comprehensively address the four-dimensional aspects of the 3D particle geometry. Two particle groups are numerically developed as below to demonstrate the efficacy of the proposed approach:

(a) Mixed morphology group, where a total of 100 polyhedral particles are modeled with a variety of morphology such that the evaluated M ranges between 1 and 3. In Figure 6, the particle models are presented with the computed Sphericity, Roundness, and M. The particle size ranges from 3 mm to 9 mm, which is also presented in the figure. The sizes are deliberately controlled such that smaller particles tend to have a more irregular morphology to distinguish the distribution from the group (b) below. It is worthwhile to note that the relation between size and morphology in these synthetic particle models is just for demonstration purpose, which does not necessarily imply smaller mineral particles in nature have a more irregular morphology.

(b) Near-sphere group, where another 100 polyhedral particles are modeled to be near-spherical. The particle models are shown in Figure 7, for which the Sphericity, Roundness, and M also are evaluated. The particle shapes in this group are all similar, so the evaluated M is near-uniformly distributed and is close to 1. The sizes are randomly selected between 3 mm and 9 mm, which are also shown in the figure.

A 2D image of each particle is used to analyze the conventional Sphericity and Roundness, for which the orientation with the maximum 2D projection area is selected as it is the most possible orientation when the particle is placed on a flat surface due to the higher stability [38]. A Sphericity and Roundness analysis code by Zheng and Hryciw [12,39] is utilized for the 2D image-based morphology analysis. A particle model is selected as an example for demonstration and shown in the green circle in Figure 6, whose corresponding data points are marked in Figure 8. The size of the example particle is 5.886 mm, and the computed Regularity $\rho$ (i.e., arithmetic average of 2D Sphericity and Roundness) is 0.3955 (thus, $1/\rho = 2.528$).

The parameter M measures the overall 3D morphology, while the conventional Sphericity and Roundness quantify the 2D morphology at both the large and intermediate scales respectively. While direct comparison may not be made as these are defined in different dimensions, the M tends to increase with elongation and angularity (i.e., lower Sphericity and lower Roundness) and vice versa. For example, the particle with a relatively high M value (= 2.313) located at 6–1 (row

# – column #) in Figure 6 has a low Sphericity and a low Roundness. On the other hand, the particle with a low M value (= 1.064) located at 1–10 in the same figure has a high Sphericity and a high Roundness.

There are particle models in Figure 6 and 7 that have practically the same M values although Sphericity and Roundness are different. For example, the particles located at 1–1, 1–2, and 1–3 in Figure 6 have practically same M values with the particles in Figure 7 located at 6–6, 2–3, and 6–5 respectively. The three particle models in Figure 6 are more equidimensional and angular (i.e., higher Sphericity and lower Roundness) compared to the particle models in Figure 7 that are more elongated and round (i.e., lower Sphericity and higher Roundness). The compensating effect results in the same M value that may be seen as a limitation, but which is necessarily not. This may be better explained with analogy to Regularity $\rho$: Regularity is a function of Sphericity and Roundness, so it is necessary to evaluate Sphericity and Roundness first (i.e., individual morphology parameters) to estimate the Regularity. It is worthwhile to note that previous studies have reported Regularity was a reasonable indicator of various soil mechanical properties at both small and large strain levels [3,4] and also facilitated the data interpretation as it is a single measure of overall morphology. Although the compensating effect may result in same $\rho$ value, the individual contributions of Sphericity and Roundness can be analyzed as the data is already available. Likewise, M is a function of A, V, and L, which can be considered as individual morphology parameters. The A, V, and L need to be evaluated first to estimate M. Although the compensating effect may result in same M value, analogous to Sphericity and Roundness to the Regularity, the individual contributions of the three parameters (i.e., A, V, and L) to M can be assessed. For example, although particle 1–2 in Figure 6 has the same M value as that of particle 2–3 in Figure 7, these two particles have different A, V, and L values from each other: the A, V, and L values of particle 1–2 (Figure 6) are 214.14 mm$^2$, 280.71 mm$^3$, and 8.719 mm, respectively, which are different from those of particle 2–3 (Figure 7), 178.87 mm$^2$, 221.99 mm$^3$, and 8.262 mm.

It is also interesting to see the differences in Roundness between these particles are larger than those in Sphericity, e.g., the Roundness values of 1–2 in Figure 6 and 2–3 in Figure 7 are 0.321 and 0.858, while the Sphericity values are 0.977 and 0.879. As discussed in Section 1, the local angularity characterizes the morphology feature that is smaller than the global form by one order of magnitude. Therefore, M appears to reasonably reflect the different contributions of morphology factors at different length scales. It however does not mean M underestimates the local angularity, as the M value can be significantly high with the local angularity only. For example, the great stellated dodecahedron in Figure 4 is not elongated but angular, and its M value is 5.33, which is very high.

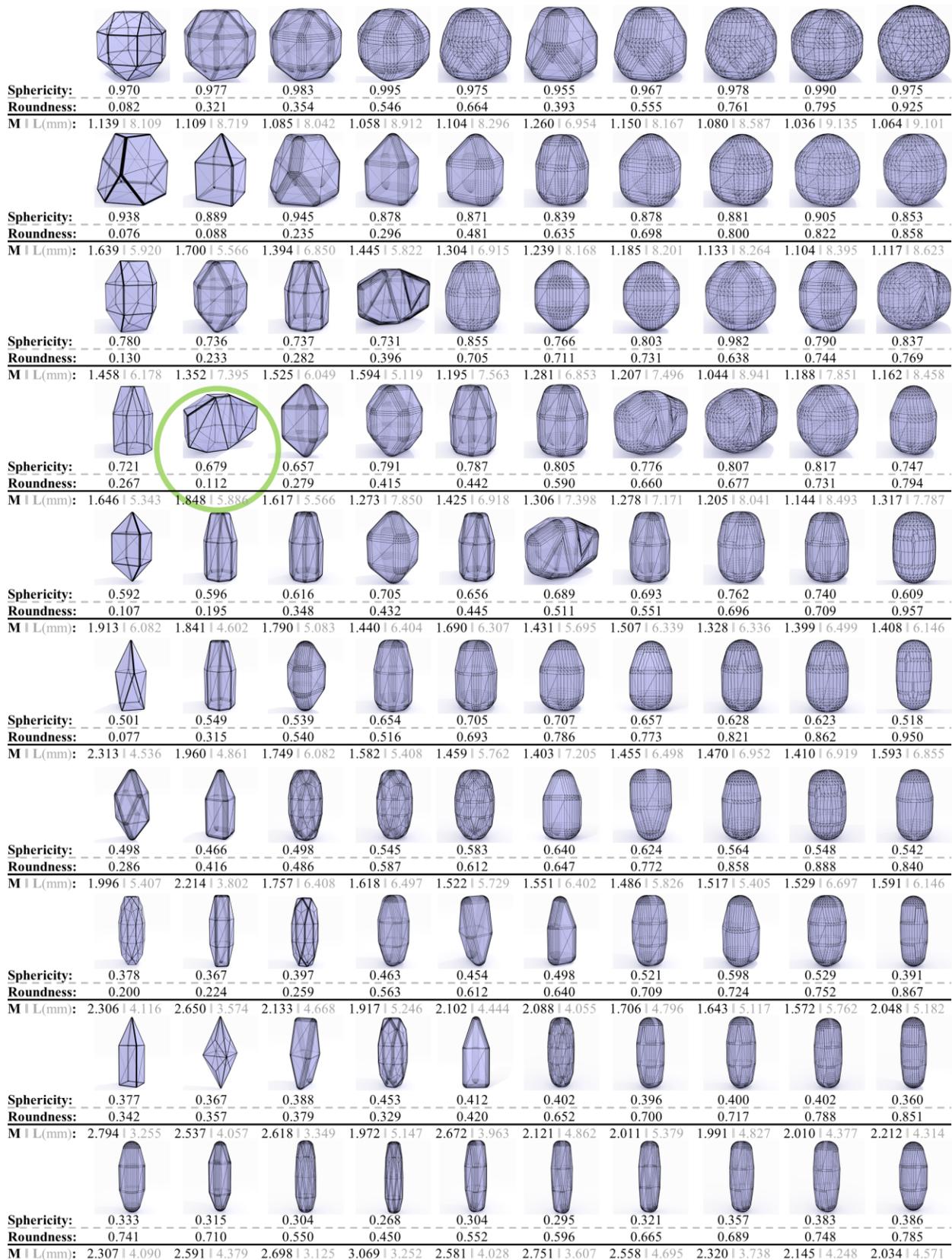

Figure 6. Particle models of mixed morphology group evaluated with Sphericity, Roundness, and proposed M values. The size L is shown next to M. The particle in the green circle is selected as an example particle and corresponding data points are shown in Figure 8.

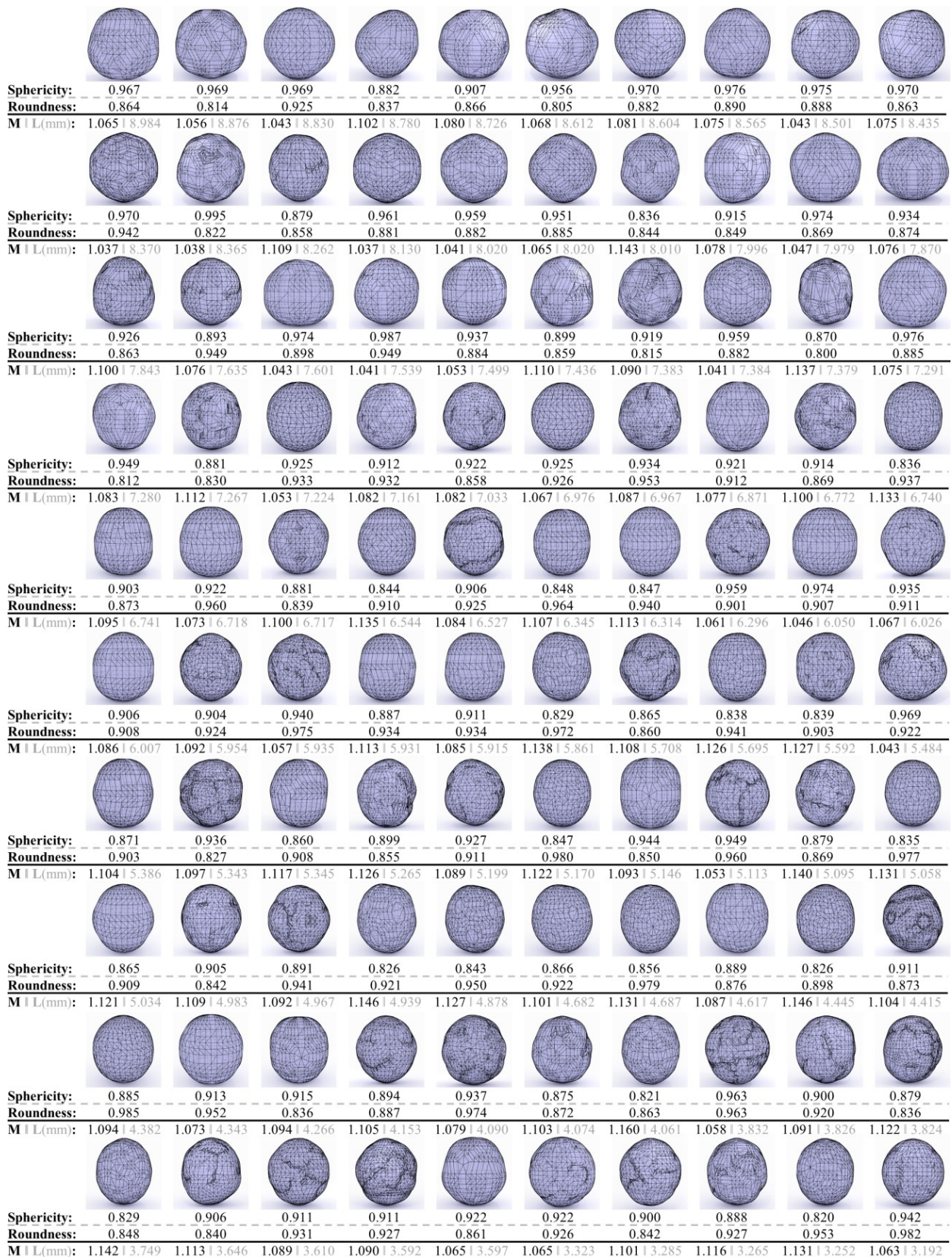

Figure 7. Particle models of near-sphere group evaluated with Sphericity, Roundness, and proposed M values. The size L is also presented next to M.

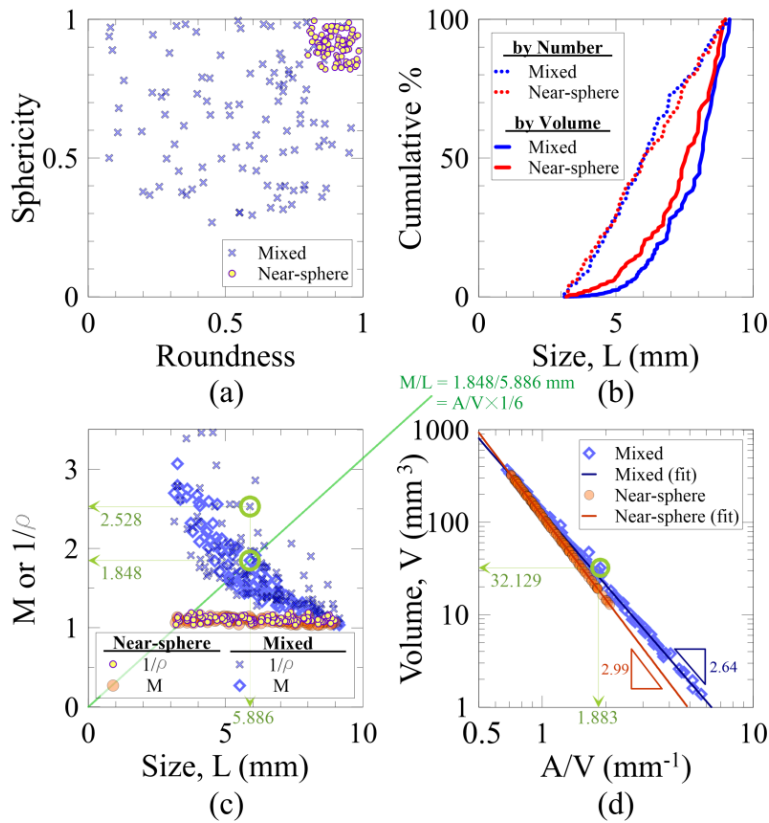

Figure 8. Evaluated particle geometry parameters for the two synthetic particle groups in Figure 6 and 7. The example particle's parameter values (selected in Figure 6) are marked with the green circles; (a) Morphology distributions characterized using conventional Sphericity and Roundness; (b) Particle size distributions evaluated in terms of particle number and volume, respectively; (c) Combined description of M and L compared to the inverse of Regularity $1/\rho$ and L. The M/L represents A/V×1/6; (d) Combined description of A/V ratio and V. The plots in Figure 8c and d can be cross-referenced using A/V = M/L×6, i.e., the A/V value in Figure 8d can be obtained from M/L in Figure 8c. The volume V of the example particle can be found in Figure 8d, from which the surface area A can be retrieved by A/V×V.

The Sphericity and Roundness of the two synthetic particle groups are plotted in Figure 8a. The data points for the mixed morphology group are spread over the space due to the variety of morphology, while those of the near-sphere group are concentrated in the upper-right corner due to the overall narrow range of morphology values with near-equidimensional and round shapes. Figure 8b shows the particle size distributions of the two groups that are evaluated by both the number and volume of particles in terms of the cumulative percentage. The size distributions by the number of particles are overall close to each other. However, the difference is apparent if evaluated by the volume because the particles in near-sphere group tend to have higher volume compared to the irregularly shaped particle of same size in the other group. Therefore, the cumulative percentage is higher for the near-sphere group. This can be explained by the example in Figure 5, where the volume of sphere in Figure 5c is larger compared to the irregularly shaped particle of the same size in Figure 5b.

The relations of geometry parameters in the distributions can be effectively described by using two 2D plots: (i) a plot presenting L vs. M (Figure 8c). This plot can relate M and L with A/V, because M/L = A/V×1/6; (ii) another plot presenting A/V vs. V (Figure 8d), where the surface area A can be retrieved by A/V×V. Therefore, M-A-V-L approach can be used to comprehensively describe the distributions of the four 3D particle geometry parameters, whereby the relations of parameters are graphically contained in these two 2D plots.

The L vs. M distributions are shown in Figure 8c for both mixed morphology and near-sphere groups. The morphology is also evaluated by Regularity $\rho$. The inverse of Regularity ($1/\rho$) is used for consistent comparison against M because a higher $\rho$ represents near-equidimensional and round shape, while a higher M represents the opposite. Figure 8c is a way that can combine the information from Figure 8a and b that are conventional characterization approach that presents morphology and size distributions separately, which therefore makes hard to examine the relation of those. On the other hand, Figure 8c combines both information in a single plot which facilitates the data interpretation whereby the relation between morphology and size in the mixed morphology group is clearly shown, i.e., the smaller the size is, the more irregular the morphology is. Despite the similar trend for both plots using M and $1/\rho$, there is more data scatter in the $1/\rho$ plot. This scatter is possibly due to the inaccuracy associated with the 2D characterization of 3D particle morphology, and the characteristic of the reciprocal function $1/\rho$, as $1/\rho$ nonlinearly increases as $\rho$ gets smaller. Figure 8c also depicts the overall similar particle morphology distributions of the near-sphere group with M~1 and $1/\rho$~1, i.e., close to the minimum possible value for both descriptors. The plots shown in Figure 8c may be represented by probabilistic density distributions if the Z-axis is used to indicate the density of data.

The A/V vs. V distributions are shown in Figure 8d. The data in Figure 8c can be cross-referenced to the A/V ratio in Figure 8d as A/V = M/L×6. For example, M/L = 1.848/5.886 mm for the example particle from Figure 8c, thus the A/V ratio can be computed to 1.883 mm$^{-1}$ (= M/L×6). The volume V of the example particle then can be found in Figure 8d, which is 32.129 mm$^3$. The surface area A then can be estimated by A/V×V, which is 60.5 mm$^2$ (=1.883 mm$^{-1}$×32.129 mm$^3$). Therefore, once two plots that represent (i) L vs. M and (ii) A/V vs. V distributions are provided, the four-geometry information (M, A, V, and L) can be retrieved for every particle. Regularity $\rho$ also represents the single measure of morphology like M. However, if $\rho$ is used in the place of M, it is not possible to cross-reference the data between the (i) L vs. 1/$\rho$ and (ii) A/V vs. V plots, so retrieval of the complete four geometry information is not feasible.

## 3.2 Recommendations for Future Research – Power Law Relation between A/V and V

Interestingly, the A/V and V in Figure 8d show a linear relation in log-log scale, which can be approximated as a 'power law.' The relation of A/V and V for the mixed morphology group can be approximated to V = (A/V)$^{-2.64}$×129.66, and then to log(V) = -2.64×log(A/V) + log(129.66). The fitted line is shown with a slope of -2.64 in the log-log space, and the equation can be reformulated to log(V) = 1.61×log(A) – 1.29, which directly relates V with A. Using A/V = M/L×6 relation, the power function can be also formulated to log(V) = -2.64×log(M) + 2.64×log(L) + 0.06, which interrelates V, M, and L. Similarly, the data for the near-sphere group can be also fitted to V = (A/V)$^{-2.99}$×116.26 and can be further formulated to find the other relations.

Considering the general relation between V and A for any geometry can be expressed as V = A$^{3/2}$×$\lambda$, where $\lambda$ is a geometry constant, the relation between V and A/V can be formulated to V = (A/V)$^{-3}$×$\beta$, where $\beta$ = 1/$\lambda^2$. Therefore, the power value of the A/V is invariably -3 for a group of particles with an identical morphology. For example, the power value is identically -3 for each group of spheres, cubes, tetrahedra or whatever morphology as far as all particles in the group have a same look. (The particles in the group, of course, will be of different sizes, otherwise the A/V and V relation will be shown as a single data point in the plot). If the relation between V and A/V of particles in a group can be generalized to $V = (A/V)^{\alpha} \times \beta$, deviation of the $\alpha$ value from -3 can indicate the degree of morphological heterogeneity of a given particle group. For example, the $\alpha$ value of near-sphere group is -2.99, which is practically -3, because the morphology in that group is nearly identical. On the other hand, the $\alpha$ value of the mixed morphology group is -2.64, and the farther deviation from -3 indicates a broader spectrum of morphology in the particle group.

The $\beta$ values can be analytically defined for some known geometries, e.g., $V = (A/V)^{-3} \times 36\pi$ for sphere, where the $\beta$ value is $36\pi$ (~113.10). The $\beta$ values of icosahedron, cube, and tetrahedron are 136.46, 216, and 374.12 respectively, which increases with geometric angularity. Therefore, the $\beta$ value (i.e., the intercept in the log-log space) may imply a characteristic of the representative morphology for a given particle group. It is noted that the $\beta$ value of near-sphere group is 116.26 which is close to that of a group of spheres (i.e., 113.10), and the $\beta$ value of mixed morphology group is 129.66, which is higher than that of the near-sphere group. While further investigation of these $\alpha$ and $\beta$ values defined in the A/V and V space is beyond the scope of this study and left for future research, the $\alpha$ value may imply the morphological heterogeneity and the $\beta$ value may give a hint of the representative morphology in particle group, which will provide another interesting interpretation to the characterization of geometry parameter distribution.

## 4 EXPERIMENTAL STUDY

Experimental study is performed to demonstrate the predictive capability of the proposed approach to estimate the influence of particle geometry on the mechanical behavior of granular materials. A set of 3D particle models are developed and then 3D printed for use in the direct shear test. The Krumbein and Sloss chart (Figure 9a) is referenced as the particle image library that is evaluated in terms of Sphericity and Roundness. The far different four morphologies at the corners in the chart, i.e., 1-1, 1-5, 4-1, and 4-5, are selected for the 3D particle development. The Fourier descriptor-based modeling technique is used to generate a realistic 3D particle model from the 2D cross-sectional images [40]. The developed 3D models are shown in Figure 9b with the evaluated M for each model. The particle 4-1 is the most irregular among the four morphologies, and therefore has the highest M value of 2.71. The irregularity evaluated by M (3D descriptor) is in order of 4-1 > 4-5 > 1-1 > 1-5 (i.e., M = 2.71 > 1.96 > 1.36 > 1.08). Regularity $\rho$ (2D descriptor) is also evaluated, and $1/\rho$ is shown in the figure. The irregularity evaluated by $1/\rho$ is in order of 4-1 > 1-1 > 4-5 > 1-5. Both M and $1/\rho$ estimate the particle 4-1 is the most irregular, and 1-5 is the opposite. However, the order of 1-1 and 4-5 is different. It appears that $1/\rho$ estimates the contribution of the particle 1-1's angularity (at the intermediate scale) is higher than the particle 4-5's elongation (at the global scale), while M estimates the contribution of the particle 4-5's elongation is higher. The Krumbein and Sloss chart represents a spectrum of mineral particle morphologies. Therefore, it is anticipated that the upper bound of M value for the mineral particles is about 3, which indicates a highly irregular particle morphology that may exist in nature.

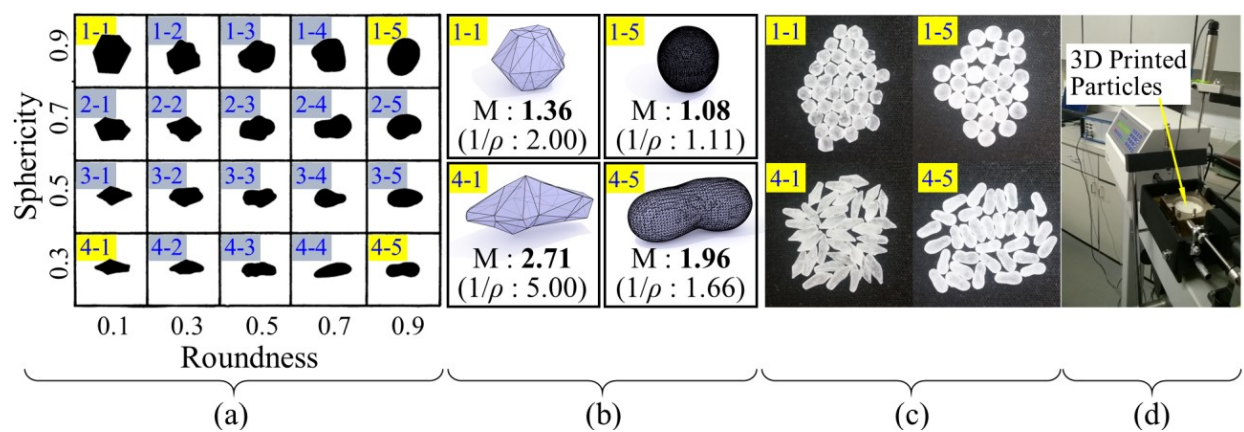

Figure 9. Development of 3D particle models and 3D printed particles; (a) Representative 2D morphology of typical mineral particles evaluated in terms of Sphericity and Roundness, modified from Krumbein and Sloss [37]. Particles are numbered in 'row # – column #' format for convenience; (b) Developed 3D particle models and the morphology quantified using M = A/V×L/6, and the inverse of Regularity, $1/\rho$; (c) 3D printed particles of the four models; (d) Direct shear test setup.

The developed 3D particle models are then 3D printed (Figure 9c) for the laboratory direct shear test (Figure 9d). Form 1+ Stereolithography (SLA) printer is used for the 3D printing [41]. Particle is printed in 25 microns of layer thickness. Therefore, the surface texture of printed particles is controlled as the particles are printed using the same material at the same printing resolution. The compressive strength and the elastic modulus of the printed objects utilizing the Formlabs resin are reported in Watters and Bernhardt [42] and Zguris [43], which are roughly comparable to those of Florida limestone [44]. The cylindrical shear box size is 63.5 mm (2.5 in) in diameter × 37 mm (1.5 in) high. Each particle model is scaled to have the same volume ($11.67 \text{ mm}^3$) such that the same number of particles can be considered per test specimen.

Each specimen of particles 1-1, 1-5, 4-1, and 4-5 is uniformly graded, i.e., composed of identical particles of same morphology, volume, surface area, and size. The initial void ratio is controlled to about 0.73 for the testing. The evaluated geometry of all four specimens are shown in Figure 10, where the evaluated $1/\rho$ is also plotted for comparison. As the four particle models are controlled to have the same volume ($11.67 \text{ mm}^3$), it is obvious that the particle size increases with the morphological elongation and angularity. Consequently, the near-spherical particle 1-5 is the smallest as 3.05 mm, while the most irregular particle 4-1 is the largest as 5.97 mm. Therefore, both M and L increase in the order of 1-5, 1-1, 4-5, and 4-1 as shown in Figure 10a. A similar tendency is shown for $1/\rho$ except for the particle 4-5 that is out of the trend, i.e., $1/\rho$ decreases while L increases. Figure 10b depicts the combined description of A/V and V, where the same trend is observed in order of 4-1 > 4-5 > 1-1 > 1-5 from the largest A/V. Likewise, the same order is estimated for the surface area A because all particles have the same V. Considering the trend in the evaluated M and the other geometry parameters in order of 4-1 > 4-5 > 1-1 > 1-5, the same trend of the mechanical performance such as shear strength and modulus is anticipated from the laboratory test.

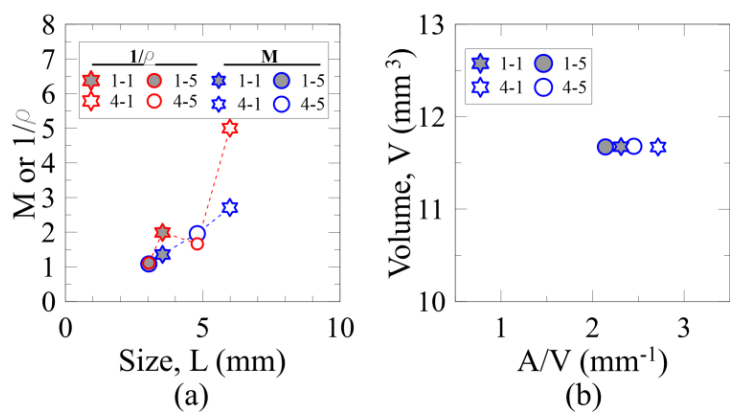

Figure 10. Evaluated geometry parameters for the four particle specimens; (a) Description of morphology and size using M and L as well as $1/\rho$ and L; (b) Description of A/V ratio and V.

Table 1. Ranges of void ratio for each specimen and the relative densities Dr for the void ratio e = 0.73

| Particle # | $e_{min}$ | $e_{max}$ | Dr (%) | Particle # | $e_{min}$ | $e_{max}$ | Dr (%) |
|---|---|---|---|---|---|---|---|
| 1-1 | 0.58 | 0.73 | ~0.00 | 1-5 | 0.62 | 0.76 | 21.42 |
| 4-1 | 0.70 | 0.87 | 82.35 | 4-5 | 0.59 | 0.73 | ~0.00 |

Table 1 shows the maximum and minimum void ratios that can be obtained for the four specimens. The particles are poured in a circular motion using a funnel onto the shear box, and tamped in layers if necessary, to achieve the initial void ratio through trial and error process. The deposition of particles is randomly done, so the specimens' fabric is not controlled. The corresponding relative densities for the initial void ratio (= 0.73) are also summarized in the table. While the initial void ratios of all specimens are controlled to be the same, particle 4-1 is relatively dense, and the other particles are relatively loose. The adopted particle sizes conform to the requirement of ASTM D3080/D3080M [45]. The specimens are tested at four different normal stresses, 40.5, 102.5, 164.4, and 226.4 kPa. The rate of shear is maintained at 1 mm/min.

Each test is repeated three times, from which the average response is obtained to eliminate the influence of initial particle arrangements [10,46]. No significant particle breakage or deformation is found at the end of tests, but some particle corners are chipped off. The specimens are manually inspected after each test, and a maximum of five particles are replaced with new ones. Figure 11a-h show the average responses of shear stress and vertical displacement at the four different normal stresses. While limited stick-slip fluctuation is shown in the test, the observed mechanical behavior is overall consistent in order of 4-1 > 4-5 > 1-1 > 1-5 from the highest to the lowest shear strength and modulus as shown in Figure 11a-d. The friction angles evaluated from the shear stress responses are shown in Figure 12, which clearly shows the strengths in the order.

A similar trend is shown for the vertical displacement in Figure 11e-h with the specimen of 4-1 showing the highest rate of dilation. The nature of specimen 4-1's vertical displacement is also different from the other specimens: the vertical displacement of 4-1 increases continuously while the responses of other specimens are relatively flattened out. The difference in the displacements is possibly attributed to the different relative densities of the specimens as shown in Table 1. A higher range of void ratio is typically obtained with a higher particle irregularity as broadly evidenced in the literature [3,47,48]. Despite the initial void ratio in all specimens controlled to 0.73, it is observed the relative density of specimen 4-1 is about 82% (i.e. relatively dense), while those of the other specimens are less than 21% (i.e., relatively loose). Therefore, the specimen 4-1 is able to continuously dilate while being sheared. The dilation angles are shown in Figure 13, which are computed based on the vertical and horizontal incremental displacements at the peak strength state. A trend of higher dilation is also shown in the same order of 4-1 > 4-5 > 1-1 > 1-5.

The test result demonstrates that the proposed approach can reasonably relate the particle scale information to the macroscopic mechanical property by its order and corroborates its predictive capability to estimate the particle geometry effect. All particles used in this study pass through 4 mm sieve and are retained in 2 mm sieve. The conventional Unified Soil Classification System classifies the coarse-grained particles in terms of particle size using the sieves, thereby the specimens in this study are classified as a same soil despite the significant differences shown in the test results. Therefore, this research finding also indicates the current specification remains to be significantly improved for enhanced classification.

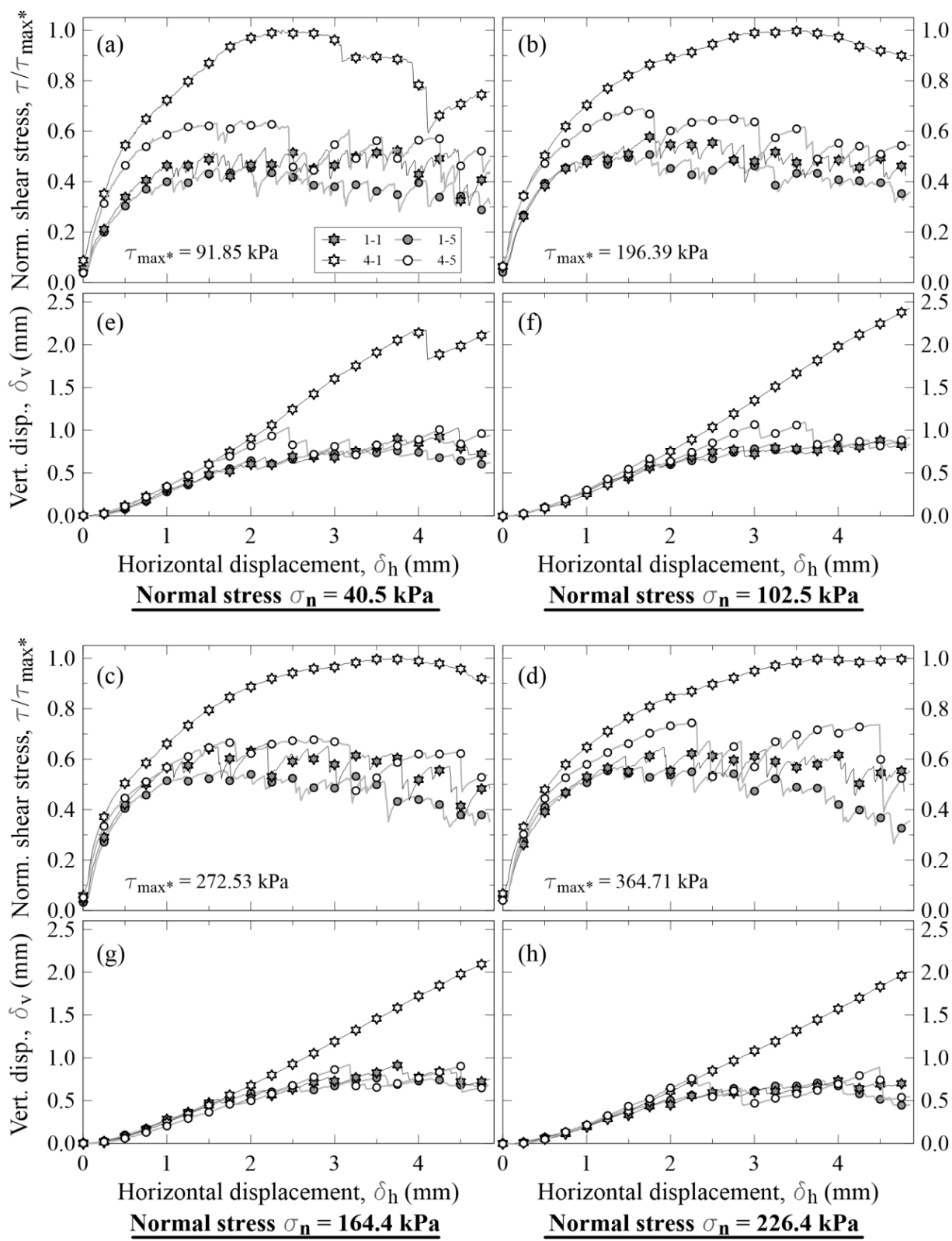

Figure 11. Direct shear test result: stress-strain curves and vertical displacements obtained at four different normal stresses (40.5, 102.5, 164.4, and 226.4 kPa); where the stress-strain curve in (a) to (d) is normalized by $\tau_{max*}$, maximum shear stress obtained from the specimen 4-1.

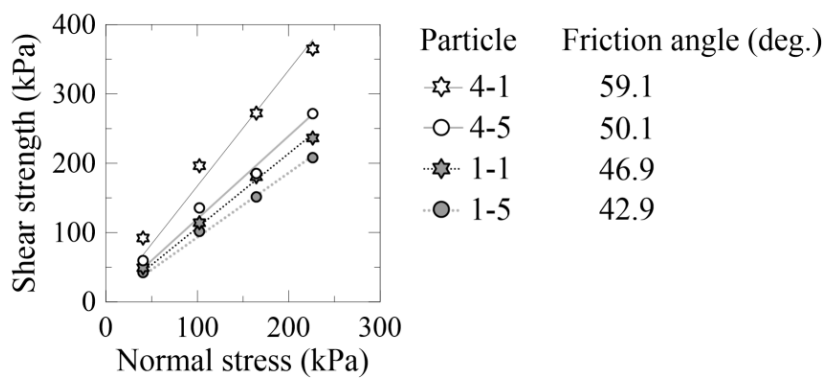

Figure 12. Friction angles evaluated from the fitting lines of shear strengths.

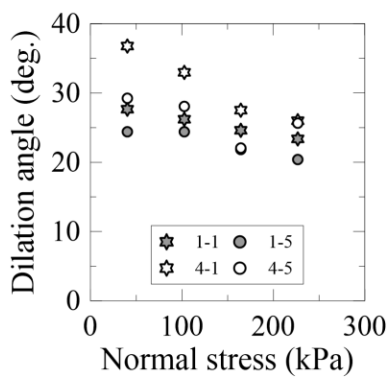

Figure 13. Dilation angles evaluated at the peak strength state.

## 5    CONCLUDING REMARKS

This study proposes a new formula, M = A/V×L/6, that interprets 3D particle morphology M as a function of the other geometry parameters, i.e., surface area A, volume V, and size L. Therefore, this MArVeLously simple formula clearly unravels the interrelation of M-A-V-L by considering A, V, and L as a set of 'morphology parameters.' This approach enables to conduct straightforward 3D particle morphology characterization using the simple formula. If one knows A, V, and L, then morphology can be evaluated in terms of M, which is more convenient than conventional approach that separately evaluated the morphology in addition to the information of A, V, and L. With the new formula, it is possible to comprehensively describe the four-dimensional aspects of the 3D particle geometry parameters in two 2D plot spaces where the individual contributions of A, V, and L (as the morphology parameters) to M can be retrieved. Therefore, this approach is complementary to the conventional approach as it enables to consider the four geometry parameters' combined influence on the behavior of granular materials. This study will contribute to enhance the predictive capability, effective aggregate quality control and proactive planning for the optimal maintenance in the field of transportation geotechnics.

A challenge of applying the proposed approach to the engineering practice may be concerned with measuring the surface area of mineral grains. This measurement can be performed by using the 3D optical characterization techniques such as photogrammetry. Notwithstanding such techniques are computationally expensive at the moment, the proposed approach is sustainable in the sense that the 3D optical characterization techniques become more accessible and affordable with newer generations of imaging equipment and the advances in image processing algorithms.

# 6 ACKNOWLEDGEMENTS

This work was sponsored in part by Florida International University (FIU) through the new faculty start-up fund given to the corresponding author, Dr. Seung Jae Lee. The opinions, findings, or conclusions expressed in this article are solely those of the authors and do not represent the opinion of FIU. The authors would like to thank Dr. Junxing Zheng and Dr. Roman Hryciw for their image-based Sphericity and Roundness analysis code used in this study. The authors appreciate Mr. Zoubair Douite's assistance to 3D-print the particles for the experimental study. Dr. Beichuan Yan's review and comments on the first draft are also appreciated. The authors would like to extend the appreciation to the editor and the anonymous reviewers for the valuable comments that helped enhance the final quality of this paper.

LIST OF CAPTIONS FOR ALL FIGURES

Figure 1. Interrelated 3D particle geometry parameters; (a) Two different particle morphology with the same size, consequently the particles have different volumes and surface areas. The size is measured in terms of the diameter of particle's bounding sphere (shown as red circles); (b) Two different particle morphology with the same volume, thus the particles have different sizes and surface areas; (c) Two different particles with same morphology having different volumes, surface areas, and sizes.

Figure 2. Conventional morphology characterization at three different scales [11].

Figure 3. Separate description of particle size and morphology distributions.

Figure 4. Morphology related to surface-area-to-volume (A/V) ratio; All particles have the same volumes of a unity; The subscript $s$ represents sphere, i.e., $A_s$ and $V_s$ are surface area and volume of the sphere. Therefore, $(A/V) / (A_s/V_s)$ indicates the A/V ratio of a particle relative to the A/V ratio of the sphere with the same volume; *The length of three edges of the stretched tetrahedron is 3.2 each, and the length of the other edges is 1.5.

Figure 5. 3D morphology (M) translated as A/V×L/6; the size is measured in terms of the diameter of the bounding sphere (shown as red circles).

Figure 6. Particle models of mixed morphology group evaluated with Sphericity, Roundness, and proposed M values. The size L is shown next to M. The particle in the green circle is selected as an example particle and corresponding data points are shown in Figure 8.

Figure 7. Particle models of near-sphere group evaluated with Sphericity, Roundness, and proposed M values. The size L is also presented next to M.

Figure 8. Evaluated particle geometry parameters for the two synthetic particle groups in Figure 6 and 7. The example particle's parameter values (selected in Figure 6) are marked with the green

circles; (a) Morphology distributions characterized using conventional Sphericity and Roundness; (b) Particle size distributions evaluated in terms of particle number and volume, respectively; (c) Combined description of M and L compared to the inverse of Regularity $1/\rho$ and L. The M/L represents A/V×1/6; (d) Combined description of A/V ratio and V. The plots in Figure 8c and d can be cross-referenced using A/V = M/L×6, i.e., the A/V value in Figure 8d can be obtained from M/L in Figure 8c. The volume V of the example particle can be found in Figure 8d, from which the surface area A can be retrieved by A/V×V.

Figure 9. Development of 3D particle models and 3D printed particles; (a) Representative 2D morphology of typical mineral particles evaluated in terms of Sphericity and Roundness, modified from Krumbein and Sloss [37]. Particles are numbered in 'row # – column #' format for convenience; (b) Developed 3D particle models and the morphology quantified using M = A/V×L/6, and the inverse of Regularity, $1/\rho$; (c) 3D printed particles of the four models; (d) Direct shear test setup.

Figure 10. Evaluated geometry parameters for the four particle specimens; (a) Description of morphology and size using M and L as well as $1/\rho$ and L; (b) Description of A/V ratio and V.

Figure 11. Direct shear test result: stress-strain curves and vertical displacements obtained at four different normal stresses (40.5, 102.5, 164.4, and 226.4 kPa); where the stress-strain curve in (a) to (d) is normalized by $\tau_{max*}$, maximum shear stress obtained from the specimen 4-1.

Figure 12. Friction angles evaluated from the fitting lines of shear strengths.

Figure 13. Dilation angles evaluated at the peak strength state.